\documentclass[prd,twocolumn,a4paper,superscriptaddress,floatfix]{revtex4}
\usepackage{graphicx,amssymb}
\usepackage[utf8]{inputenc}
\usepackage[T1]{fontenc}
\usepackage{colortbl}
\usepackage{xcolor}

\begin{document}

%My commands
\newcommand{\be}{\begin{equation}}
\newcommand{\ee}{\end{equation}}
\newcommand{\bq}{\begin{eqnarray}}
\newcommand{\eq}{\end{eqnarray}}
\newcommand{\bsq}{\begin{subequations}}
\newcommand{\esq}{\end{subequations}}
\newcommand{\bc}{\begin{center}}
\newcommand{\ec}{\end{center}}
\newcommand {\R}{{\mathcal R}}
\newcommand{\al}{\alpha}
\newcommand\lsim{\mathrel{\rlap{\lower4pt\hbox{\hskip1pt$\sim$}}
    \raise1pt\hbox{$<$}}}
\newcommand\gsim{\mathrel{\rlap{\lower4pt\hbox{\hskip1pt$\sim$}}
    \raise1pt\hbox{$>$}}}

\title{Antipredator behavior in the rock-paper-scissors model}

\author{J. Menezes}  
\email[Electronic address: ]{jmenezes@ect.ufrn.br} 
\affiliation{Escola de Ci\^encias e Tecnologia, Universidade Federal do Rio Grande do Norte\\
Caixa Postal 1524, 59072-970, Natal, RN, Brazil}
\affiliation{Institute for Biodiversity and Ecosystem Dynamics, University of Amsterdam, Science Park 904, 1098 XH Amsterdam, The Netherlands}

\date{\today}
\begin{abstract}
When faced with an imminent risk of predation, many animals react to escape consumption. \mbox{Antipredator} strategies are performed by individuals acting as a group to intimidate predators and minimize the damage when attacked. We study the antipredator prey response in spatial tritrophic systems with cyclic species dominance using the rock-paper-scissors game. The impact of the antipredator behavior is local, with the predation probability reducing exponentially with the number of preys in the predator's neighborhood. In contrast to the standard Lotka-Volterra \mbox{implementation} of the rock-paper-scissors model, where no spiral waves appear, our outcomes show that the antipredator behavior leads to spiral patterns from random initial conditions. The results show that the predation risk decreases exponentially with the level of antipredator strength. Finally, we investigate the coexistence probability and verify that antipredator behavior may jeopardize biodiversity for high mobility. Our findings may help biologists to understand ecosystems formed by species whose individuals behave strategically to resist predation.
\end{abstract}
\maketitle

\section{Introduction}

The spatial segregation of species is a fundamental \mbox{issue} in ecology\cite{ecology}. To this purpose, many authors have conducted experimental and theoretical studies to understand how interactions among individuals are responsible for ecosystem formation and stability
\cite{Nature-bio, BUCHHOLZ2007401}. In this scenario, the experiments with bacteria \textit{Escherichia coli} unveiled the role of space to preserve biodiversity \cite{Coli}. There is a cyclic dominance among three bacteria strains that can be described by the rock-paper-scissors game rules \cite{bacteria}. However, the cyclic dominance is not sufficient to guarantee coexistence, but individuals must interact locally. The consequence is the formation of spatial domains occupied mostly by individuals of the same species \cite{Allelopathy}.
The same phenomenon is observed in groups of lizards \cite{lizards} and coral reefs \cite{Extra1}. 

Given the relevance of the cyclic dominance in maintaining biodiversity, 
stochastic simulations of the rock-paper-scissors model have been an essential tool to comprehend how spatial patterns appear and affect species persistence \cite{Szolnoki_2020, Szolnoki-JRSI-11-0735}. 
The simulations may be realized either considering a conservation law for the total number of individuals (Lotka-Volterra implementation \cite{doi:10.1021/ja01453a010,Volterra}) or with the presence of a variable density of empty spaces (May-Leonard implementation \cite{Reichenbach-N-448-1046,Avelino-PRE-86-036112,uneven}). This paper focuses on the Lotka-Volterra version, where random mobility competes with local predation interaction. In this case, the spiral patterns observed in the May-Leonard stochastic simulations of the rock-paper-scissors models are not present, as shown in Ref.~\cite{PhysRevE.78.031906} (see also \cite{Nagatani2018,PhysRevE.89.042710,Pereira,PhysRevE.99.052310} for generalizations of the rock-paper-scissors game).

It is well known that behavioral strategies play a vital role in evolutionary biology \cite{BUCHHOLZ2007401}. For example, movement strategies drive individuals either searching for natural resources (see \cite{Motivation1,butterfly}) or seeking refuges against enemies \cite{refuge1,refuge2}. Recently, it has been shown that \mbox{behavioral} movement tactics may give advantages to species that compete for space in cyclic models \cite{Moura}.
Another well-known animal behavior is the resistance against predation \cite{DefenseAnimals}. For example, many vertebrates and invertebrates perform Thanatonis (death feigning) tactics to inhibit predator attack \cite{Thanatonis,AntiFish1}. As an antipredator strategy, prey mites \textit{Tetranychus urticae} emits an odor when exposed to the predatory mite \textit{\mbox{Phytoseiulus} \mbox{persimilis}} to reduce the oviposition, and the consequent predator population growth \cite{ContraAtacck2}. It has also been reported in Ref.~\cite{ContraAtacck1} that 
the western flower thrips \textit{Frankliniella \mbox{occidentalis}} can kill the eggs of their predator, the predatory mite \textit{Iphiseius degenerans}.
To protect themselves against predators, spider mites also vary the nest size, and web density \cite{NinhosdeMites2,MitesWeb1,MitesWeb2}. Furthermore, antipredator behavior leads individuals to form groups \cite{Grouping2}. Lizards \textit{Lampropholis delicata} live together
to join efforts to respond to predation threat \cite{LizardB1}. The alarm vocalizations alerting against the presence of predators is one of the benefits observed in groups of California bighorn sheeps \textit{Ovis canadensis californiana} \cite{Grouping1}. In addition, experiments have shown that the antipredator behavior is crucial to stabilize the predator-prey system at a population level \cite{NinhosdeMites1}. Studies of the effects of grouping on predator-prey interactions in cyclic models are scarce in the literature. Recently, some authors presented the results for a generalization for four species \cite{Cazaubiel}. They investigated the system's stability when both predators and preys congregate to maximize their performance in the game.

In this paper, we investigate the role of the antipredator behavior in cyclic nonhierarchical tritrophic systems. We consider that i) individuals of all species have the same ability to respond to predation when threatened; ii) the efficiency of the antipredator response depends on the prey group size. Our goal is to understand how the local dynamics of predator-prey interactions change population growth and biodiversity. To this purpose, we consider the Lotka-Volterra implementation of the rock-paper-scissors game, where interactions are predation and mobility. We introduce a local effect on predation, reducing the predation probability as a function of the prey group size. This means that each predator has an effective predation probability which depends on its local reality - the number of preys in the neighborhood. We also consider an antipredator strength factor to model the responsiveness to predators. 
The outline of this paper is as follows. In Sec. II, we introduce the model describing how the stochastic rules are implemented and how antipredator behavior is modeled. In Sec. III, we show the effects of the antipredator behavior on the spatial patterns, comparing the results with the standard Lotka-Volterra implementation of the rock-paper-scissors game. In Sec. IV, we investigate the dynamics of the spatial densities in the presence of the antipredator response. We quantify the spatial patterns using the autocorrelation function for various levels of antipredator strength in Sec. V. The impact of the antipredator behavior on an individual's predation risk is presented in Sec. VI, while the biodiversity maintenance in terms of the individual's mobility is addressed in Sec. VII.
Finally, our comments and conclusions appear in Sec. VIII.

\label{Introduction}
%=======================================================================================================

\section{The Model}

\begin{figure}[t]
\centering
%%% Please, do not change the scale %%%
\includegraphics[width=40mm]{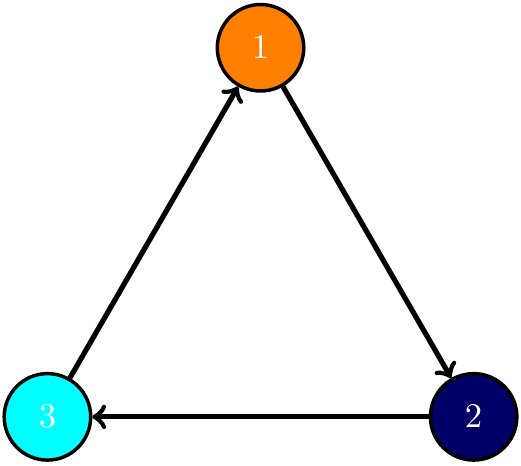}
\caption{Illustration of the predation interaction rules among species in the rock-paper-scissors model.}
	\label{fig1}
\end{figure}

We consider a system composed of three species that dominate each other according to the popular rock-paper-scissors game rules, as illustrated in Fig. \ref{fig1} - the arrows indicate a cyclic trophic dominance among the species. 
The different species are labeled by $i$ (or $j$) with $i,j= 1,...,3$, with the cyclic identification $i=i\,+\,3\,\alpha$ where $\alpha$ is an integer. Accordingly, individuals of species $i$ prey individuals of species $i+1$.
The dynamics of individuals' spatial organization happen in a square lattice with periodic boundary conditions. We follow the Lotka-Volterra numerical implementation, where the total number of individuals is conserved  \cite{doi:10.1021/ja01453a010,Volterra}. In this scenario, the total number of individuals is always equal to $\mathcal{N}$, the total number of grid points - each grid point contains one individual.
The possible interactions are:
\begin{itemize}
\item 
Predation: $ i\ j \to i\ i\,$, with $ j = i+1$. Every time one predation interaction occurs,  
the grid point occupied by the individual of species $i+1$ is occupied by a offspring of species $i$.
\item
Mobility: $ i\ \odot \to \odot\ i\,$, where $\odot$ means an individual of any species. When moving, an individual of species $i$ switches positions with another individual of any species.
\end{itemize}
 
This work considers that individuals of every species perform
antipredator behavior: predation is harmed by a defensive response of the prey group surrounding the predator. The collective antipredator action leads to a decrease in predation probability that depends on the group size and the preys' resistance level. 
To implement the model, we first define an antipredator effect range, $R$, as the maximum distance that one prey can interfere with the predator action, which is measured in units of the lattice spacing. For a predator of species $i$, the effective predation probability is a function of the fraction of individuals of species $i+1$ within a disk of radius $R$, centered at the predator. Defining that $\mathcal{G}_{max}$ is the maximum group size - the number of individuals that fit within a disk of radius $R$ - and considering that $\mathcal{G}$ is the actual group size surrounding the prey, the effective predation probability is given by
\be
p_{eff}\,=\,p\,e^{-\kappa\,\frac{\mathcal{G}}{\mathcal{G}_{max}}},
\ee
where $\kappa$ is the antipredator strength factor, a real parameter defined as $\kappa\geq0$, indicating how the preys' opposition jeopardizes predation.
$\kappa=0$ represents the standard model, where predation probability is given by $p_{eff}=p$. When preys fully compose the predator's neighborhood, $\mathcal{G\,}=\,\mathcal{G}_{max}$, predation probability is minimal: $p_{eff}=p\,e^{-\kappa}$. Conversely, when one prey is alone, it does not have the help of its co-specifics to react to the predator. In this case, the likelihood of being consumed is maximal: $p_{eff}=p\,e^{-\kappa/\mathcal{G}_{max}}$.

We assumed random initial conditions, where each grid point is given an individual of an arbitrary species. Initially, the total numbers of individuals of every species are the same: $I_i=\mathcal{N}/3$, for $i=1,2,3$. The interactions were implemented by assuming the Moore neighborhood, i.e., individuals may interact with one of their eight immediate neighbors. The simulation algorithm follows three steps: i.) sorting a random individual to be the active one; ii.) drawing one of its eight neighbor sites to be the passive individual; iii.) randomly choosing an interaction to be executed by the active individual ($m$ and $p_{eff}$ are the mobility and predation probabilities).
If the active and passive individuals (steps i and ii) allow the raffled interaction (step iii) to be performed, one timestep is counted. Otherwise, the three steps are redone.
Our time unit is called generation, which is the necessary time to $\mathcal{N}$ interactions to occur. Throughout this paper, we present results obtained for $R = 3$, which means that the prey group size surrounding a predator varies in the range $ 0 \leq \mathcal{G} \leq 28 $.

% fig 1 %%%%%%%%%%%%%%%%%%%%%
% fig 1 %%%%%%%%%%%%%%%%%%%%%
%%%%%%%%%%%

%%%%%
%=======================================================================================================
\section{Spatial Patterns}
%================================================================================================================

We begin the numerical study by performing a single simulation for the cases $\kappa=0$ (the standard model), $\kappa=1$, $\kappa=3$, and $\kappa=5$. The realizations run in square lattices with $600^2$ sites for a timespan of $5000$ generations, assuming $p=2/3$ and $m=1/3$. The dynamics of the spatial patterns are shown in videos in Refs.~\cite{video1,video2,video3,video4}. The snapshots depicted in the upper left, upper right, lower left, and lower right panels show the spatial configuration at the end of the simulations for $\kappa=0$, $\kappa=1$, $\kappa=3$, and $\kappa=5$, respectively. The colors follow the scheme in Fig.~\ref{fig1}, where orange, dark blue, and cyan dots show individuals of species $1$, $2$, and $3$, respectively.

\begin{figure}
	\centering
	\includegraphics*[width=4.2cm]{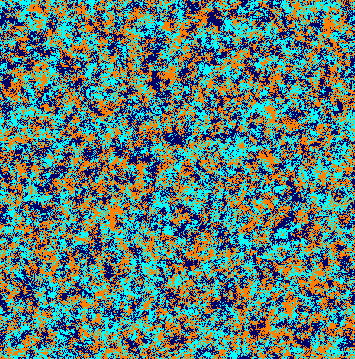}
	\includegraphics*[width=4.2cm]{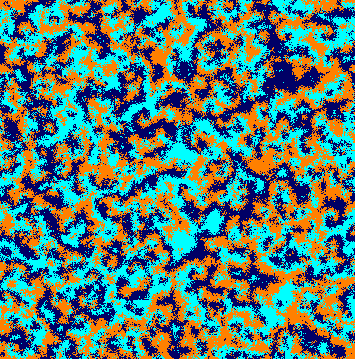}
	\includegraphics*[width=4.2cm]{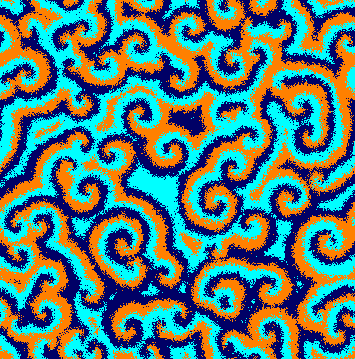}
	\includegraphics*[width=4.2cm]{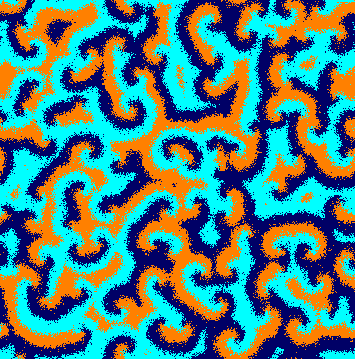}
\caption{Snapshots of simulations of the rock-paper-scissors game illustrated in Fig. \ref{fig1} running in square lattices with $600^2$ grid points. Each dot shows an individual according to the color scheme in Fig. \ref{fig1}. All simulations started from the same random initial conditions. The upper left, upper right, lower left, and lower right panels show the results for $\kappa=0$ (the standard model), $\kappa=1$, $\kappa=3$, and $\kappa=5$.
}
 \label{fig2}
\end{figure}

The aleatory distribution of individuals leads to a
high predation rate in the initial stage of the simulation. In the standard model ($\kappa=0$), predators find and consume preys everywhere without resistance. The local species segregation continuously changes because of the cyclic predation interactions. For example, when a group of individuals of species $1$ appears, it is consumed and substituted by individuals of species $2$. The new spatial domain of species $2$ is, in its turn, destroyed by individuals of species $3$, that serve as food for species $1$. The consequence is the formation of irregular groups during the simulation, as shown in the upper left panel of Fig.~\ref{fig2}. The video \cite{video1} shows the dynamics of the spatial patterns for the standard model during the entire simulation.

When antipredator behavior is considered, predation is no longer as likely probable to any predator. There is a local effect on the predation probability: the larger the prey group size, the more difficult it is to eat the prey. Because of this, predators that are close to conspecifics have more chances of devouring the prey. This accounts for the growth of the species spatial domains, leading to spiral patterns. As one sees in the upper right, lower left, and lower right panels of Fig.~\ref{fig2}, $\kappa$ influences the spatial pattern formation. The more intense the prey's resistance, the less likely predation to occur away from the boundaries of predator-dominated domains. Moreover, for higher $\kappa$, fewer predation interactions happen, increasing the effective mobility rate, and consequently, the spatial domain size (\cite{Avelino-PRE-86-036112}). Videos \cite{video2} ($\kappa=1$), \cite{video3} ($\kappa=3$), and \cite{video4} ($\kappa=5$) show the dynamics of the species spatial segregation.

%%%%%%%

%=======================================================================================================
\section{Dynamics of Species Densities}
%================================================================================================================

To quantify the population dynamics, we computed the
spatial density $\rho$, defined as the fraction of the grid occupied by individuals of one species. Due to the cyclic tritrophic chain's symmetry - inherent to the rock-paper-scissors model - the average spatial densities are the same for each species. Therefore, we concentrate only on the spatial density of species $1$, that is function of time $t$, i.e., $\rho(t) = I_1(t)/\mathcal{N}$. 

\begin{figure}
	\centering
	\includegraphics*[width=8.2cm]{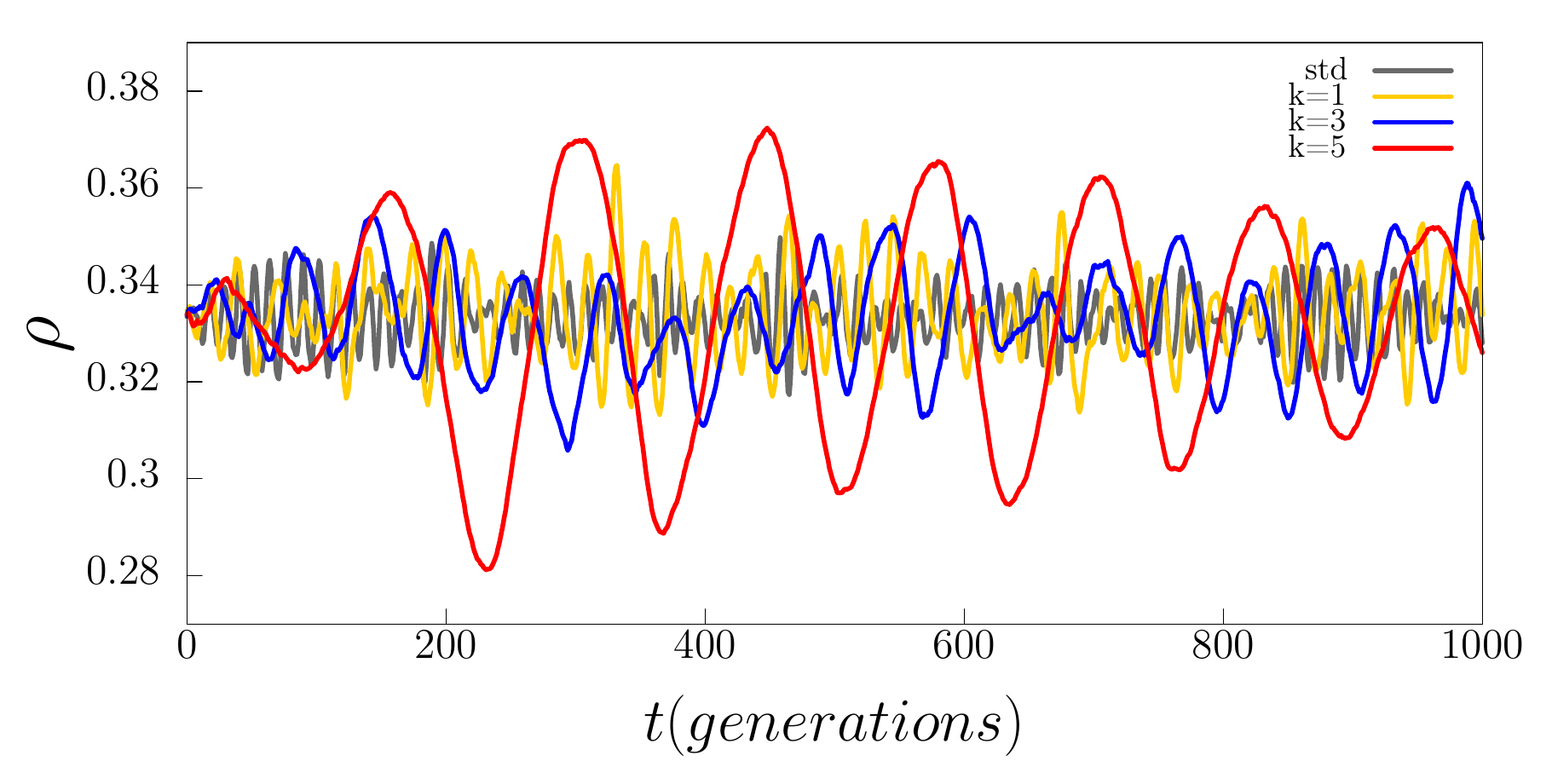}
\caption{Temporal changes of spatial species densities $\rho$ in the simulations presented in Fig.~\ref{fig2}. The grey, yellow, blue, and red lines represent the results for antipredator strength factor $\kappa=0$ (standard model), $\kappa=1$, $\kappa=3$, and $\kappa=5$, respectively. }
 \label{fig3}
\end{figure}

The temporal changes in spatial densities of the simulations showed in Fig. \ref{fig2} were depicted in Fig. \ref{fig3}.
The grey line shows the dynamics of $\rho$ for the standard model (\cite{video1}), while the yellow, blue, and red lines represent the results for antipredator strength factor $\kappa=1$ (\cite{video2}), $\kappa=3$ (\cite{video3}), and $\kappa=5$(\cite{video4}), respectively. 
The outcomes show that the territorial dominance of species $i$ ($i=1,2,3$) is cyclic, as expected in the predator-prey models \cite{doi:10.1021/ja01453a010,Volterra}. The amplitude and frequency of the spatial densities increase for larger $\kappa$, resulting from the spiral pattern formation. 

\begin{figure}
	\centering
	\includegraphics*[width=5.7cm]{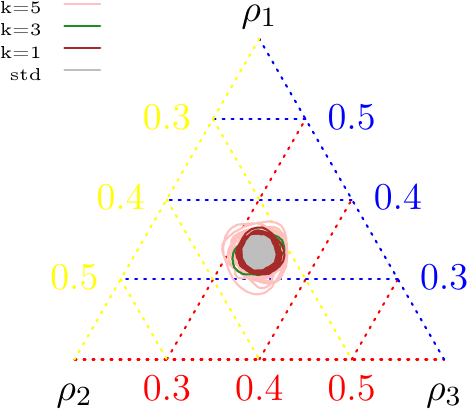}
\caption{Ternary diagram of the species densities in the tritrophic systems presented in Fig.~\ref{fig2}. The orbits for the cases $\kappa=0$, $\kappa=1$, $\kappa=3$, and $\kappa=5$, are showed by the grey, brown, green, and pink lines, respectively.}
 \label{fig4}
\end{figure}

The species densities are also depicted in a ternary diagram in Fig.~\ref{fig4} for $\kappa=0$ (grey line), $\kappa=1$ (brown line), $\kappa=3$ (green line), and $\kappa=5$ (pink line). Even though the fluctuations of the species densities increase for larger $\kappa$, species coexist because the spatial domain average size is smaller than the grid size (\cite{Reichenbach-N-448-1046,PhysRevE.97.032415}).

%%%
%%%

%=======================================================================================================
\section{Autocorrelation Function}
%=======================================================================================================
%=========

The spatial autocorrelation function quantifies the species spatial segregation, measuring how individuals of the same species are spatially correlated. Again, we focus on computing the autocorrelation function of species $1$.

%%%%
\begin{figure}[t]
	\centering
	\includegraphics*[width=8.2cm]{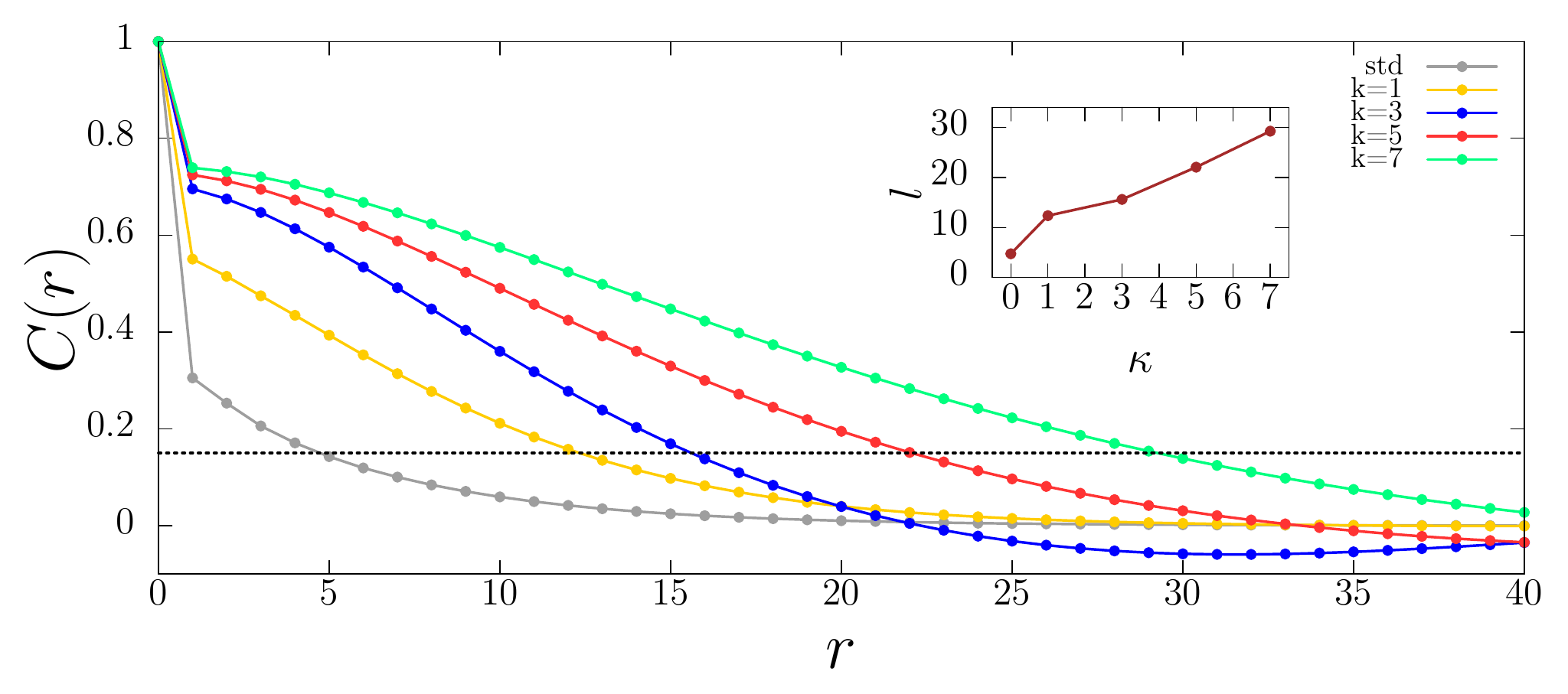}
\caption{Autocorrelation functions $C(r)$. They grey, yellow, blue, red, and green lines show the outcomes for $\kappa=0$, $\kappa=1$, $\kappa=3$, $\kappa=5$, and $\kappa=7$, respectively. The horizontal dashed black line indicates the threshold assumed to calculate the characteristic length. The inset shows the characteristic length as a
function of $\kappa$.}
 \label{fig5}
\end{figure}
The autocorrelation function is computed from the inverse Fourier transform of
the spectral density as
\be
C(\vec{r}') = \frac{\mathcal{F}^{-1}\{S(\vec{k})\}}{C(0)},
\ee
where $S(\vec{k})$ is given by
\be
S(\vec{k}) = \sum_{k_x, k_y}\,\varphi(\vec{\kappa}),
\ee
with $\varphi(\vec{\kappa})$ being the following Fourier transform
\be
\varphi(\vec{\kappa}) = \mathcal{F}\,\{\phi(\vec{r})-\langle\phi\rangle\}.
\ee 
The function $\phi(\vec{r})$ represents the spatial distribution of individuals of species $1$ ($\phi(\vec{r})=0$ and $\phi(\vec{r})=1$ indicate the absence and the presence of an individual of species $1$ in the the position $ \vec{r}$ in the lattice, respectively).

The spatial autocorrelation function is computed as
\be
C(r') = \sum_{|\vec{r}'|=x+y} \frac{C(\vec{r}')}{min (2N-(x+y+1), (x+y+1)}.
\ee
Finally, we found the spatial domains' scale for $C(l)=0.15$, where $l$ is the characteristic length.

Figure \ref{fig5} shows how the spatial autocorrelation function changes in terms of the radial coordinate $r$, for various values of $\kappa$. The results were averaged from a set of $100$ simulations with lattices with $\mathcal{N}=300^2$ - each simulation started from different random initial conditions. The spatial configuration was captured after $3000$ generations, for $p=2/3$ and $m=1/3$.
The yellow, blue, red, and green lines show the autocorrelation functions for $\kappa=1$, $\kappa=3$, $\kappa=5$, and $\kappa=7$, respectively in the upper panel. The grey line shows the autocorrelation function for the standard model, $\kappa=0$.
The horizontal black line represents the threshold considered to calculate the length scale, $C(l)\, =\, 0.15$.
The inset shows the characteristic length in terms of $\kappa$. Accordingly, in the presence of the local effects of the antipredator behavior, the spatial clustering of individuals of the same species is remarkably augmented, reflecting the visualized effects in the spatial patterns.

%=======================================================================================================
\section{Predation Risk}
%================================================================================================================
%%%
\begin{figure}
	\centering
	\includegraphics*[width=8.2cm]{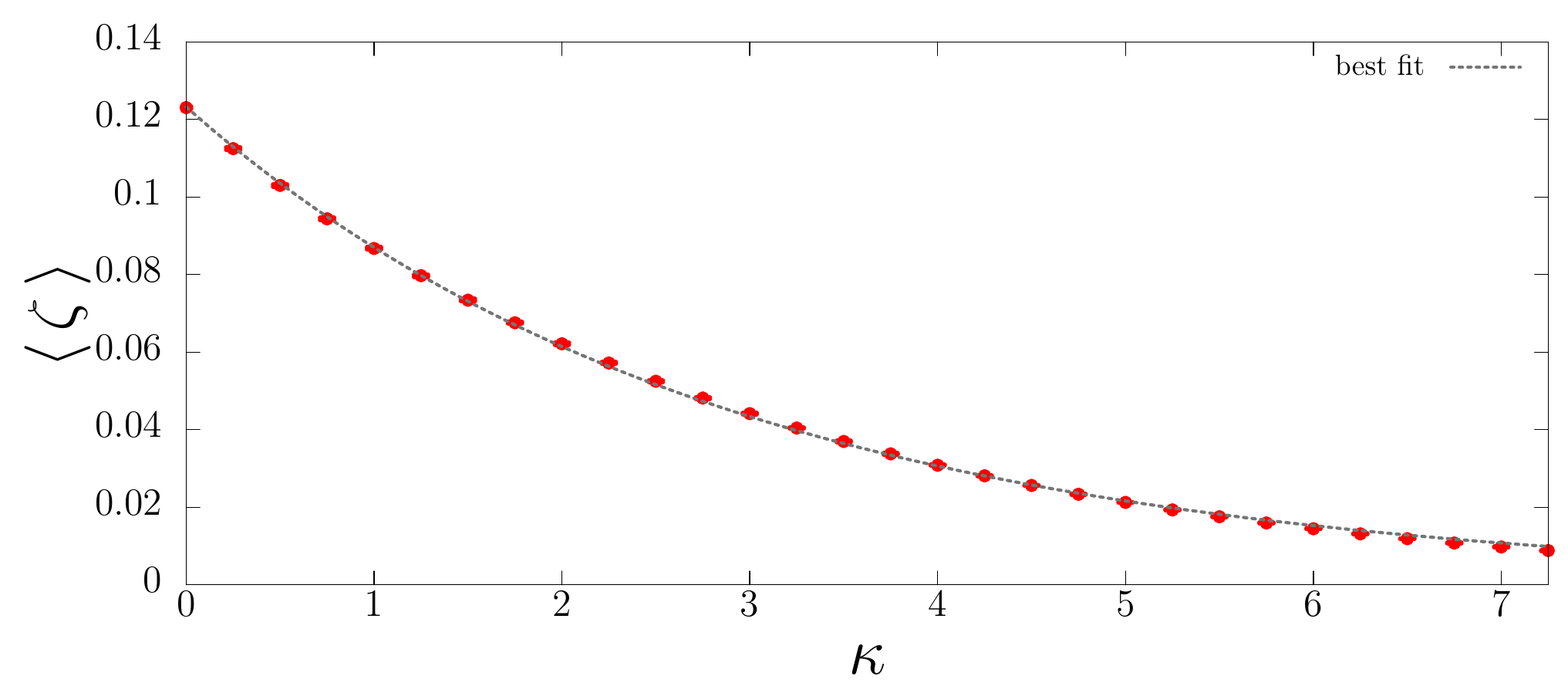}
\caption{Mean predation risk as a function of the antipredator strength factor. The results were averaged from a set of $100$ simulations of squares lattices with $300^2$ points. The error bars shows the standard deviation.}
 \label{fig6}
\end{figure}
%%%

Now, we aim to comprehend how collective antipredator behavior reduces the chances of an individual being killed. For this reason, we compute the predation risk $\zeta$. Because of the symmetry of the rock-paper-scissors model, individuals of every species have the same predation risk. We then focus on calculating $\zeta$ for species $1$.

We first count the total number of individuals of species $1$ at the beginning of each generation. Subsequently, we count how many individuals of species $1$ are consumed during the generation. The ratio between the number of preyed individuals and the initial amount is defined as the predation risk, $\zeta$. To avoid the noise inherent in the pattern formation period, we calculate the predation risk considering only the second half of the simulation. Besides, we averaged the results every $30$ generations. 

To understand how the predation risk is sensitive to the antipredator strength factor $\kappa$, we run sets of $100$ realizations with different random initial conditions for each value of $\kappa$. The mean value of the predation risk, $\langle\zeta\rangle$ is depicted in Fig.~\ref{fig6} for $0\leq\kappa\leq7.25$. The error bars show the standard deviation; $\kappa=0$ represents the standard model. We verified that the predation risk decreases exponentially when the antipredator strength factor grows. The best fit to the mean predation risk is given by 
\begin{equation}
\langle \zeta \rangle\, =\, \langle \zeta_0 \rangle\,e^{-\mu\,\kappa},
\end{equation}
where $\langle \zeta_0 \rangle = 0.123 \pm 3.7 \times 10^{-4}$ is the predation risk in the standard model, whereas $\mu = 0.35 \pm 1.6 \times 10^{-3}$. The fit shows the influence of the neighborhood on the predation risk. This means that $\mu$ computes the average antipredator effect caused by the prey group surrounding every predator.

%=======================================================================================================
\section{Coexistence Probability}
%================================================================================================================

Finally, we aim to discover how antipredator behavior jeopardizes biodiversity. 
To this purpose, we performed $1000$ simulations in lattices with $120^2$ grid points for $ 0.05\,<\,m\,<\,0.95$ in intervals of $ \Delta\, m\, =\,0.05$. The simulations started from different random initial conditions and run for a timespan of $120^2$ generations. Predation probability was set to $p\,=\,1-m$. 
Coexistence occurs if at least one individual of each species is present at the end of the simulation, $I_i (t=120^4) \neq 0$ with $i=1,2,3$. Otherwise, the simulation results in extinction. The coexistence probability is the fraction of realizations resulting in coexistence. 

\begin{figure}
	\centering
	\includegraphics*[width=8.2cm]{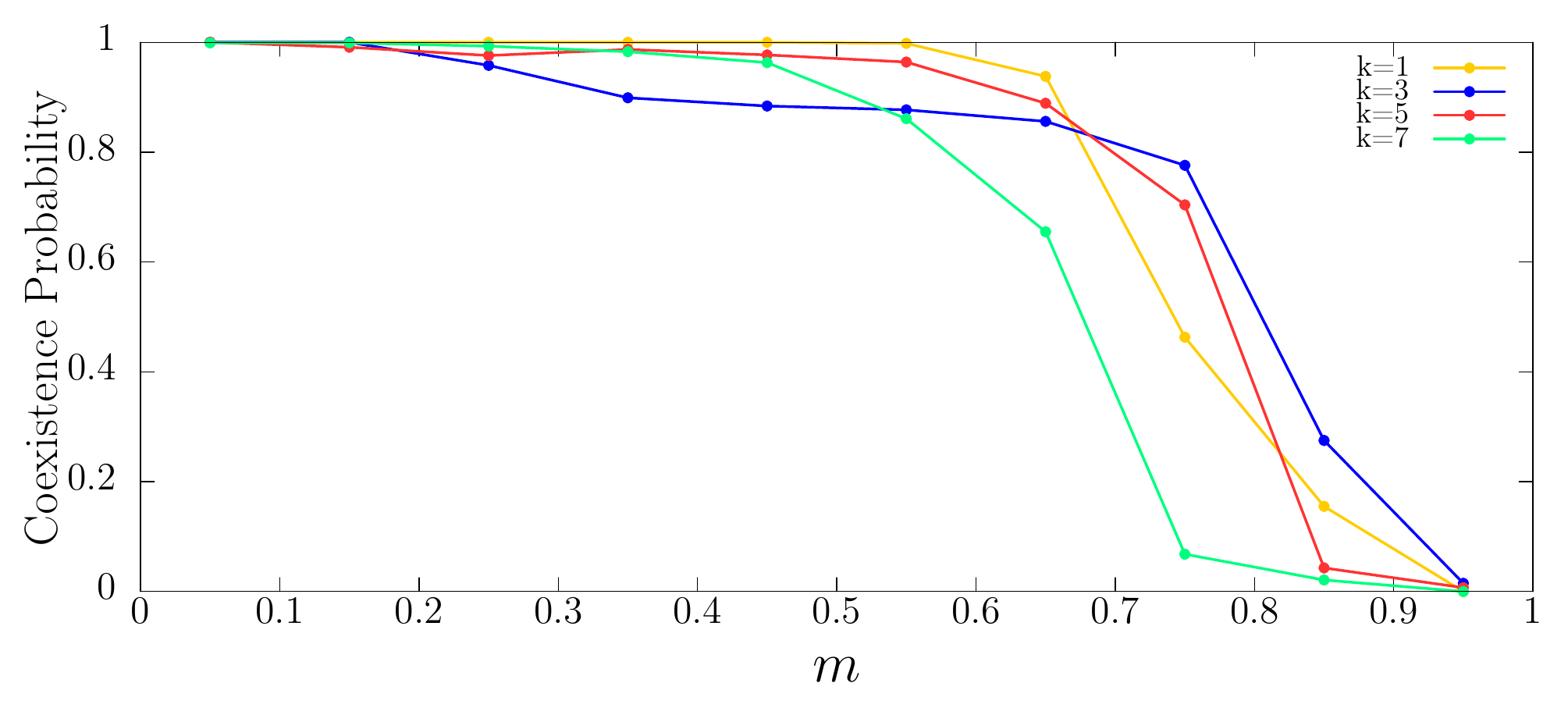}
\caption{Coexistence probability as a function of the mobility probability $m$. The yellow, blue, red, and green lines show the results for $\kappa=1$, $\kappa=3$, $\kappa=3$, and $\kappa=7$, respectively.
The results were obtained by running $1000$ simulations in lattices with $120^2$ grid points running until  $120^2$} generations.
 \label{fig7}
\end{figure}

We investigated how coexistence probability is affected by the antipredator strength factor $\kappa$. Figure ~\ref{fig7} shows the results for various values of $\kappa$. Yellow, blue, red, and green lines show the coexistence probability for $\kappa=1$, $\kappa=3$, $\kappa=5$, and $\kappa=7$, respectively. 
As Fig.~\ref{fig7} indicates, the coexistence probability does not behave monotonically. This nonlinearity is a result of both the change of the spatial patterns resulting from the increase in $\kappa$ and the stochasticity caused by the different mobility probabilities \cite{Reichenbach-N-448-1046}.

For $m<0.55$, species biodiversity is more threatened when $\kappa=3$, while $m \geq 0.55$, the chances of all species persisting is lower for $\kappa=7$.

%%%%%%%%%%%%%%%%%%%%

\section{Comments and Conclusions}
We studied a tritrophic predator-prey system described by the nonhierarchical cyclic rock-paper-scissors game. In our model, 
collective behavior is responsible for the prey group's opposition against a predator. Considering that the predation probability decreases exponentially with the prey group size and the antipredator strength, we performed a series of stochastic numerical simulations to understand the effects on the spatial pattern formation and species spatial densities. We also investigated the impact on predation risks and the coexistence probability.

Our main result shows that collective antipredator behavior leads to spiral pattern formation. Here, the Lotka-Volterra implementation of the rock-paper-scissors model shows the presence of spiral waves in on-lattice simulations. In Ref. \cite{PhysRevE.78.031906}, the authors claim that this is not possible when a conservation law for the total number of individuals is assumed, nor is it likely to form any other visible spatial pattern (this was confirmed in our simulations for $\kappa=0$). The authors also claimed that spiral patterns only appear whether the total number of individuals on the lattice is no longer conserved (\cite{PhysRevE.78.031906}), the so-called May-Leonard implementation. In this case, besides individuals of species $1$, $2$, and $3$, empty spaces are also considered on the lattice (see Refs.~\cite{Reichenbach-N-448-1046,Avelino-PRE-86-036112,uneven,PhysRevE.99.052310}).
Indeed, for the Lotka-Volterra implementation, spiral patterns were observed exclusively in off-lattice simulations (\cite{0295-5075-121-4-48003,basins,PhysRevE.82.066211}).
Here, the spiral patterns result from the influence of the group size on local antipredator behavior: individuals on the borders between predator-dominated and prey-dominated are more likely to succeed in preying. This is similar to the May-Leonard implementation, where individuals need empty spaces, mostly present on domain boundaries, to reproduce. 

Our findings also show how predation risk decreases exponentially with the antipredator strength factor. More, we verified that the collective antipredator behavior might jeopardize biodiversity for higher mobility probabilities. Our outcomes help to understand how regional conditions affect predators' performance and, consequently, change population dynamics in cyclic models. The results may also shed light on investigations of complex systems in other areas of nonlinear science.
 
\section{Acknowledgments}
We thank Beatriz Moura and Enzo Rangel for enlightening discussions. We acknowledge ECT/UFRN, CNPQ/Fapern and IBED-Universiteit van Amsterdam for financial support.

\bibliography{ref}

\end{document}